# Exchange Interactions in a Dinuclear Manganese (II) Complex with Cyanopyridine-N-oxide Bridging Ligands

A.S. Markosyan [a,b], I.Yu. Gaidukova [a], A.V. Ruchkin [a],

A.O. Anokhin [c], V.Yu. Irkhin [*c], M. V. Ryazanov [d], N.P. Kuz'mina [d], V.N. Nikiforov [a]

[a] Faculty of Physics, M.V. Lomonosov Moscow State University, 119992 Moscow, Russia

[b] Department of Applied Physics, Stanford University, USA

[c] Institute of Metal Physics, Ural Division of the Russian, Ekaterinburg, Russia

[d] Faculty of Chemistry, M.V. Lomonosov Moscow State University, 119992 Moscow, Russia

The magnetic properties of dinuclear manganese(II) complex [Mn(hfa)$_2$cpo]$_2$ (where hfa is hexafluoroacetylacetonate anion and cpo is 4-cyanopyridine-N-oxide) are presented. The non-monotonous dependence of magnetic susceptibility is explained in terms of the hierarchy of exchange parameters by using exact diagonalization. The thermodynamic behavior of pure cpo and [Mn(hfa)$_2$(cpo)]$_2$ is simulated numerically by an extrapolation to spin $S = 5/2$. The Mn-Mn exchange integral is evaluated.

## Introduction

Among a large number of molecule-based magnetic materials the heterospin systems consisting of 3d transition metal ions and organic free radicals as ligands are of special interest[1-3]. When the radical molecules have a non-zero spin, they can form various ferrimagnetic spin structures with relatively high Curie point. The indirect exchange in most of these complexes is, however, negative. Ferromagnetic interactions between Mn$^{II}$ ions were observed for several compounds only: with azido[4], diazine[5], Robson-type[6] and phthalocyaninate[7] macrocyclic ligands.

In the search and design of high-spin molecule-based magnets, a new strategy can provide polynuclear complexes due to the special geometry of the chemical bonds. Polynuclear metal-radical complexes were studied owing to their relevance to many important naturally occurring processes[8-10] and their unique magnetic properties. Some of these systems have been found to behave as single-molecule magnets[11] and further they have been shown to undergo quantum tunneling of the magnetization[4].

Some of us have reported the synthesis, crystal structure and preliminary data on magnetic behavior of the mixed ligand complex [Mn(hfac)$_2$(cpo)]$_2$ (**1**), formed by interaction of manganese(II) hexafluoroacetylacetonate dihydrate with 4-cyanopyridine-N-oxide (cpo)[12]. In this paper we present the detailed results on magnetic behavior of this dinuclear complex (**1**) and demonstrate that it turns out to be an excellent model object to describe theoretically the hierarchy of exchange interactions in molecular magnets.

Although the cpo molecule has a singlet ground state and all the interactions in **1** are negative, due to the special geometry of the magnetic clusters the Mn spins align parallel and the complex has a high-spin magnetic ground state. This results in a non-trivial temperature dependence of magnetic susceptibility. We explain this dependence on the basis of the Heisenberg model by using straightforward diagonalization.

**Experimental**

The complex [Mn(hfac)$_2$(cpo)]$_2$ (**1**) was prepared as polycrystalline powder from Mn(hfac)$_2$·2H$_2$O and cpo in chloroform by method described in[12] and characterized by data of elemental, IR and XRD analyses.

Elemental analyses (C, H, N) were performed by the Microanalytical Service of Moscow State University. The Mn content was determined by titrimetric analysis[13]. IR spectra were recorded as Nujol mulls in KBr plates on a Perkin–Elmer FTIR 1720 spectrometer in the 4000–400 cm$^{-1}$ region.

Anal. Calc. for C$_{32}$H$_{12}$N$_4$F$_{24}$O$_{10}$Mn$_2$: C, 32.6; H, 1.0; N, 4.75; Mn, 9.3. Found: C, 32.6; H, 0.8; N, 4.4; Mn, 9.1%. IR (Nujol, cm$^{-1}$): 1660s m(C=O),1500br s, 1380m, 1260s, 1220s, 1150s, 870m, 805m, 740m, 665s, 585m, 565m, 440m.

Powder XRD patterns were collected on STOE STADI/P diffractometer (Cu-K$_{\alpha 1}$, λ = 1.54060Å, curved Ge (111) monochromator) in transmission geometry at room temperature. The experimental powder XRD pattern of polycrystalline **1** was identical with XRD pattern calculated by using the data of single crystal structure determination[12].

The sample degradation was controlled by elemental and XRD analysis. The data confirmed that the composition of **1** did not changed under long-time storage in dissicator and during magnetic measurements.

Magnetic susceptibility data (2-300 K) and the magnetization curve *M(H)* at *T* = 2 K for the polycrystalline sample were collected with SQUID-based sample magnetometer on a Quantum Design MPMS instrument. The experimental data were corrected for the diamagnetism of the constituent atoms estimated from Pascal's constants.

**Results and discussion**

The crystal structure of the complex **1** has been used here for interpretation of both magnetic experiments and theoretic modeling data. This structure includes centrosymmetric discrete [Mn(hfac)$_2$(cpo)]$_2$ dinuclear entities (Figure 1), two manganese (II) ions being double-bridged by oxygen atoms of the cpo-ligands.

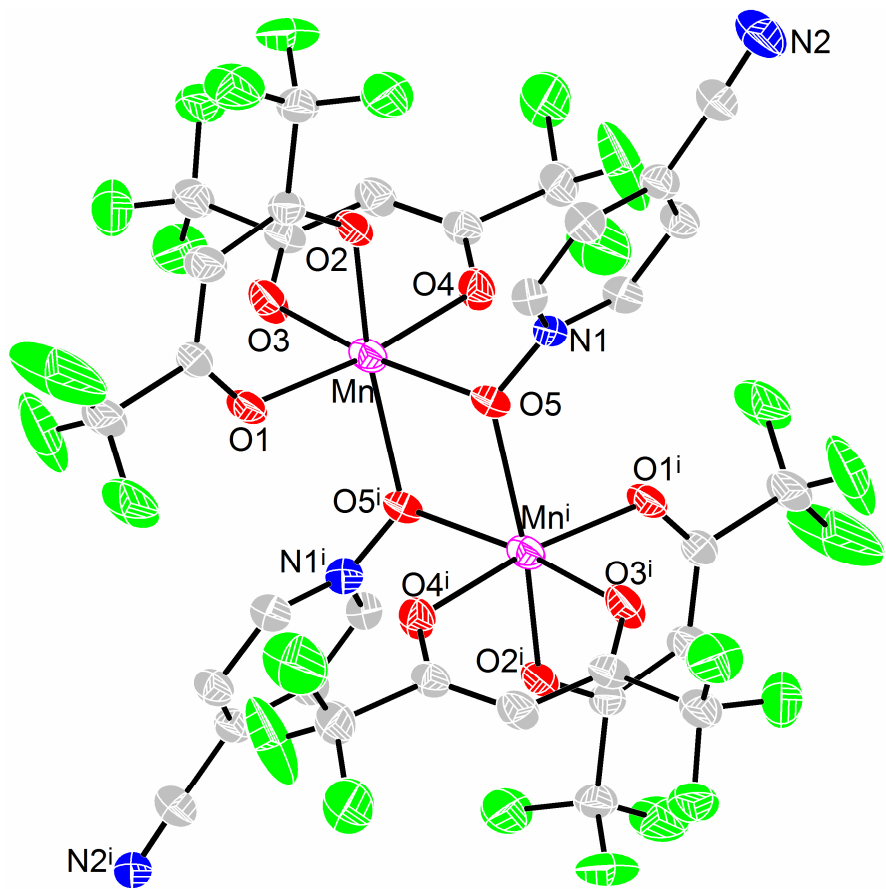

(**a**)

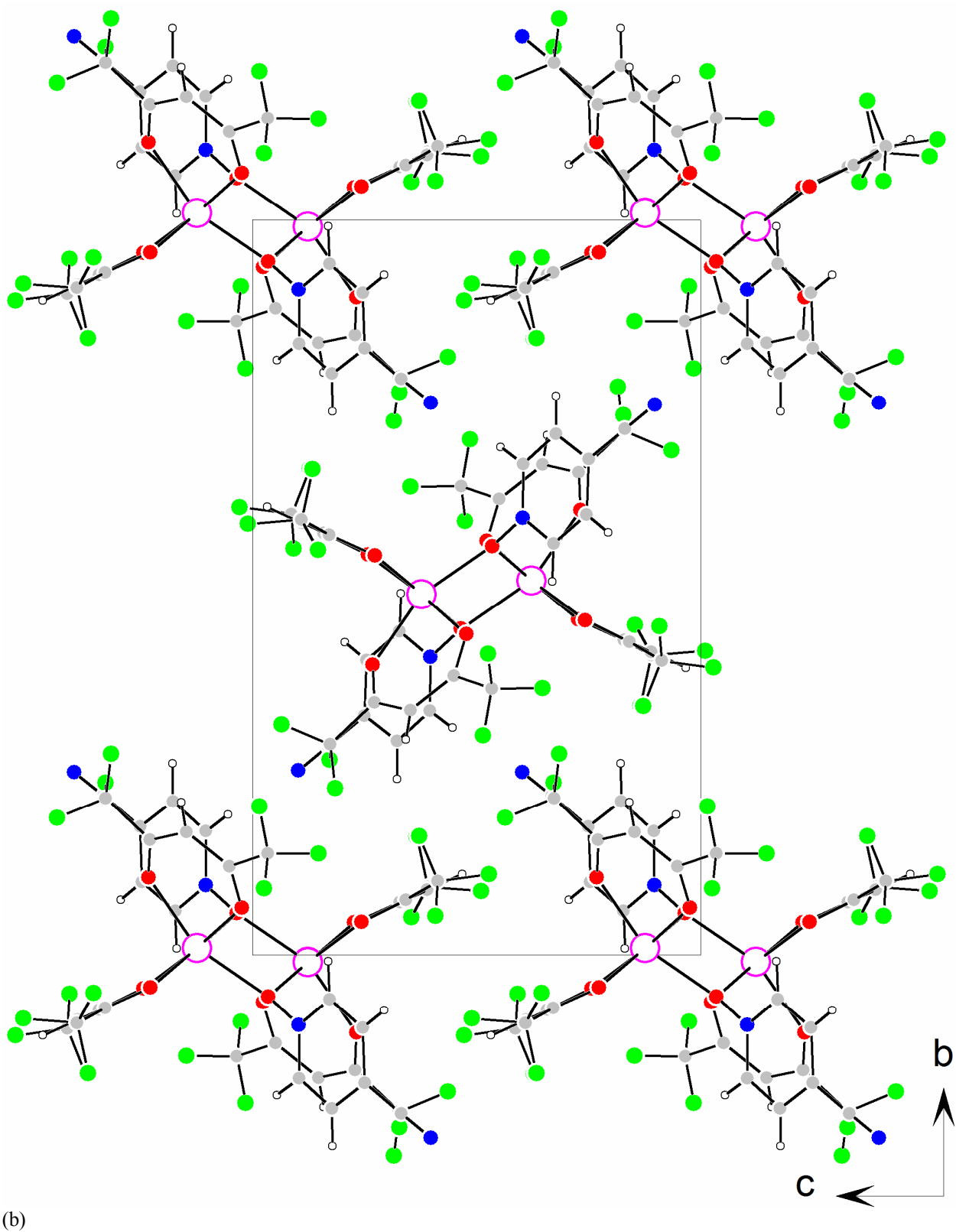

(b)

Figure 1. Molecular structure (a) and the crystal packing fragment (b) of the complex **1**. Hydrogen atoms are omitted for the sake of clarity

Due to the planar configuration of the Mn-O(5)-O(51)-Mn(l) bridging core, the complex has inversion center. The intramolecular Mn-Mn separation is equal to 3.567(1) Å, and the angle Mn-O(5)-Mn(l) is 107.00(6) deg. In this structure, each Mn ion lies in a distorted octahedral environment and is bounded to six oxygen atoms belonging to two bridging cpo ligands and two (hfac) anions, which act as bidentate terminal ligands. The cyano groups of the cpo ligands are not involved in the coordination of any metal center. In the unit cell the dinuclear molecules are well isolated from each other: the distances between the metal ions belonging to neighboring molecules are longer than 9 Å.

The clusters build blocks and are packed in a way to form a monoclinic structure. For this structure, a tetrahedral coordination is quite natural (each cluster is surrounded by 4 other clusters. However, it can also be 6 or 8. This is not forbidden, since the monoclinic structure might be a distorted derivative of a high-symmetry structure. Inside the cluster, the coordination number of Mn ions is 6; for the N-O species it is 3 and for N-C it is very low: 3 or even 2 (unfortunately we do not know how the blocks are connected with each other).

**Magnetic Properties**

The temperature-dependent magnetic properties of the complex **1** are presented in Figure 2 in the form of $\chi_m T$ ($\chi_m$ is the molar magnetic susceptibility). At low temperatures $\chi_m$ shows a sharp increase characteristic for a non-singlet magnetic ground state (with a non-zero total spin in molecules) and the value of $\chi_m T$ reaches 12.4 cm$^3$mol$^{-1}$K at 2 K. This tendency points also to absence of antiferromagnetic intermolecular exchange interactions. With increasing temperature, a steep decrease of $\chi_m$ is followed by a broad minimum located near 60 K. The high-temperature asymptotic limit of $\chi_m T$, 8.9 cm$^3$mol$^{-1}$K, is close to (though slightly higher than) the value expected for two non-interacting Mn$^{II}$ ions, 8.75 cm$^3$mol$^{-1}$K.

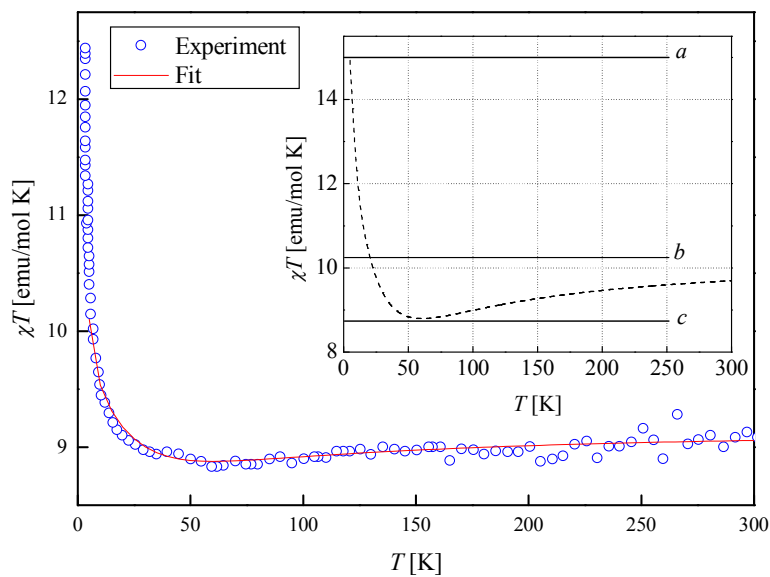

Figure 2. The temperature dependence of the product $\chi_m T$ for [Mn(hfac)$_2$(cpo)]$_2$. Experimental data are shown by open circles. The solid line represents the best fit calculated numerically using Eqs (1)–(3). The inset shows the simulated temperature variation of $\chi_m T$ for $\lambda = 0$, $P = 1$ and $C = 0$ (well-isolated clusters); the horizontal solid lines show the asymptotic values of $\chi_m T$ (see the text for details).

Our measurements for isolated polycrystalline cpo radical obtained in the powder form[5] have shown that this molecule has two $s = 1/2$ spins coupled antiferromagnetically. The space location of the magnetic orbitals that can bear these spins is unknown yet. However, we can suggest that they are located mostly at the nitrogen sites as in nitronyl nitroxide units[3]. Using this result the magnetic clusters in **1** were treated as 6-nuclear units combined from two Mn$^{II}$ spins $S = 5/2$ and four $s = 1/2$ spins of two cpo molecules, as shown in Figure 3.

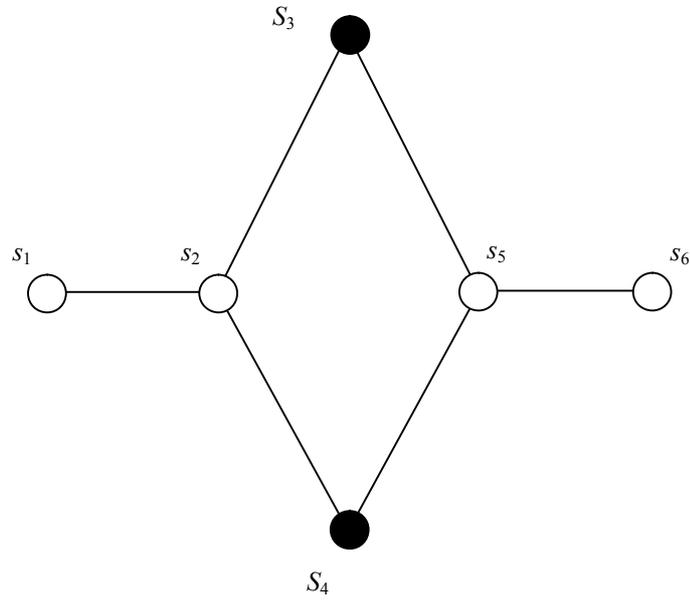

Figure 3. A schematic representation of the spin structure of magnetic clusters in the complex **1**. The black symbols represent the Mn$^{II}$ centers (with spins $S_i$) and the open symbols are the spins $s_i$ of the cpo molecules.

The Hamiltonian of this cluster within the frame of the Heisenberg model reads as

$$H = 2J_1(s_1 s_2 + s_5 s_6) + 2J_2(s_2 S_3 + s_2 S_4 + S_3 s_5 + S_4 s_5) \qquad (1)$$

where $S_i$ are the Mn$^{II}$ spins, $s_i$ are the spins located at the cpo-molecules, $J_1$ and $J_2$ are the corresponding exchange integrals. The energy spectrum of this cluster has 576 levels.

The Hamiltonian (1) was solved numerically and the eigenvalues were found by a straightforward diagonalization. The magnetic susceptibility of $N_A$ non-interacting clusters was then calculated through the partition function $Z = \sum_n \exp(-E_n/k_B T)$:

$$\chi_{m0} = N_A g \mu_B Z'_x / Z, \qquad (2)$$

where $x = g\mu_B H/k_B T$, all the other symbols having their usual meanings. The exchange integrals $J_1$ and $J_2$ were then evaluated by fitting the theoretically obtained values of $\chi_m T$ to the experimental data shown in Figure 2. In the theoretical expression, the intercluster exchange interaction was taken into account in the molecular field approximation by introducing a molecular-field coefficient $\lambda$. The final expression that was compared with the experimental data has the following form:

$$\chi_m T = P(\chi_{m0} T / (1 - \lambda \chi_{m0})) + C, \qquad (3)$$

where $P$ is the purity factor and $C$ is the Curie constant describing the impurity paramagnetic centers that could be present in the sample. The best fit was obtained with the following values of the fitting parameters: $2J_1/k_B = -22.8$ K, $2J_2/k_B = -0.76$ K, $\lambda = +0.25$ mol/emu, $P = 0.2$, and $C = 3.56$ (see Figure 2). The solution shows that there is a substantial amount of free paramagnetic centers contributing to $\chi_m T$. Assuming these are well-isolated Mn$^{II}$ centers, the obtained value of $C$ makes up about 81% of such centers present in the sample. This value is in good agreement with the purity factor $P = 0.2$.

As can be seen from Figure 2, this set of parameters describes all the details of the temperature variation of $\chi_m T$. Although at low temperatures the magnetic structure of **1** is ferromagnetic, the minimum at 60 K appears due to a delicate balance between $J_1$ and $J_2$ even at $\lambda = 0$ (see the inset in Figure 2). The low-temperature ferromagnetic state of the cluster is a consequence of a special spin geometry of the cluster that cancels the total spins of cpo in the ground state and provides ferromagnetic $Mn^{II}$-$Mn^{II}$ coupling (though both $J_1$ and $J_2$ are negative). Hence the low-temperature limit of $\chi_m T$ is 15 cm$^3$mol$^{-1}$K (horizontal line *a* in the inset). Since $J_2 \ll J_1$, with increasing temperature the cluster's $\chi_m T$ becomes a sum of those of $2N_A$ non-interacting $Mn^{II}$ centers, while the contribution from the cpo-spins is still negligible. The $\chi_m T$ value is therefore close to 8.75 cm$^3$mol$^{-1}$K (horizontal line *b* in the inset). With further increasing temperature, when the energy states of the cpo dimer become excited, $\chi_m T$ tends asymptotically to the high-temperature limit for independent two spins 5/2 and four spins 1/2, 10.25 cm$^3$mol$^{-1}$K (horizontal line *c* in the inset). Note that the minimum is very sensitive to the ratio $J_2/J_1$ and cannot be observed if $J_2$ is comparable with or exceeds $J_1$.

As it can be seen from these results, the intercluster exchange is small in the complex **1** (assuming the coordination number of cluster 4 gives the value of the intercluster exchange integral of 0.023 K), and the sample remains paramagnetic down to at least 2 K. Figure 4 shows the magnetization curve of **1** at 2 K. The magnetization curve has a Brillouin-like shape, however slightly different from that of non-interacting spins $S = 5/2$, corresponding to the magnetic ground state. For a correct modeling of the magnetization process, one has to take into account that owing to a small value of $J_2$ (0.38 K) some low-lying levels in **1**, beside the ground-state multiplet, are also populated at this temperature. For the above-evaluated exchange integrals, the energy spectrum of the cluster consists of 12 sets of closely located sublevels with different values of $S_{tot}$. While the energy gap between these sets is over 20 K, the levels within each set are separated by less than 0.5 K. The first and second sets of the energy spectrum are shown in the inset in figure 4. The theoretical curve in Figure 4 was then fitted to experiment taking into account four lowest energy levels and using the following expression for $M(H)$ derived from the quantum-mechanical definition of magnetization:

$$M = \frac{N_A g^2 \mu_B^2}{3 k_B (T-\theta)} \left[ P_1 \times \left( \frac{\sum_{n=1}^{4}(2S_n+1)\exp(-E_n/k_B(T-\theta))\coth(S_n+1/2)y}{\sum_{n=1}^{4}\exp(-E_n/k_B(T-\theta))\sinh(S_n+1/2)y} - \frac{1}{2}\coth(y/2) \right) + P_2 \times B_{5/2}(y) \right], \quad (4)$$

with $y = g\mu_B H / k_B (T-\theta)$ and $\theta = \lambda C$ being the paramagnetic Curie temperature originating for the intercluster exchange interaction. In this procedure, the energies $E_n(S_n)$ were fixed parameters equal to: $E_1(5) = 0$, $E_2(4) = 0.305$, $E_3(4) = 0.381$, and $E_4(3) = 0.686$ K. The first term in Eq. (4) is the contribution from the native phase **1**, and $B_{5/2}(y)$ describes the contribution from the impurity 5/2 spins, i.e. from non-interacting $Mn^{II}$ ions. The best fit of Eq. (4) to the experimental data was obtained with the fitting parameters $P_1 = 0.363$, $P_2 = 0.637$ and $\theta = 0.45$ K. This fit corresponds to 36 % decay of molecular Mn complex.

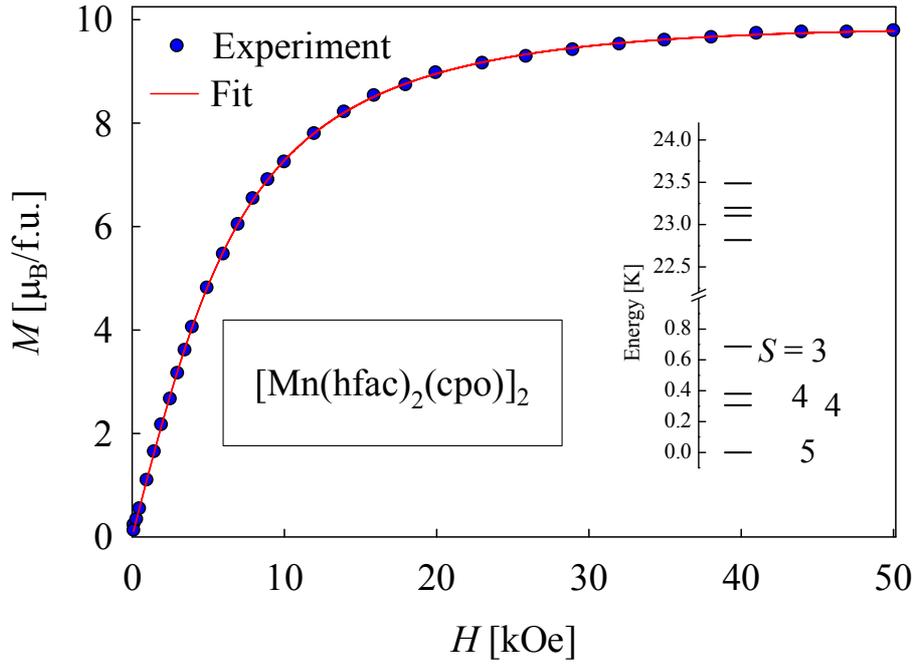

Figure 4. The field dependence of the magnetization of **1** at 2 K (full circles). The solid line corresponds to Eq. (4) (see text) with the energy spectrum evaluated from the $\chi_m T$ vs. $T$ dependence. The inset show two lowest quartets of the energy spectrum at zero magnetic field. For four lowest-lying multiplets contributing to the magnetization process at 2 K, the spin values are given by respective numbers.

We present also the experimental data on a pure cpo sample (Figs. 5,6).

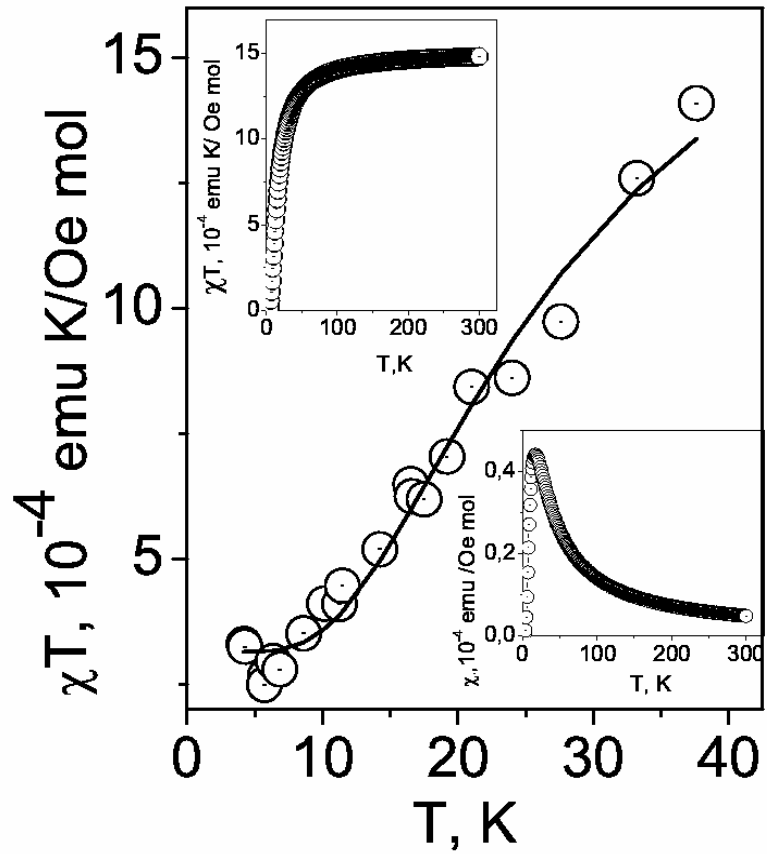

Figure 5. Magnetic susceptibility of pure cpo. Insets show the high-temperature behavior

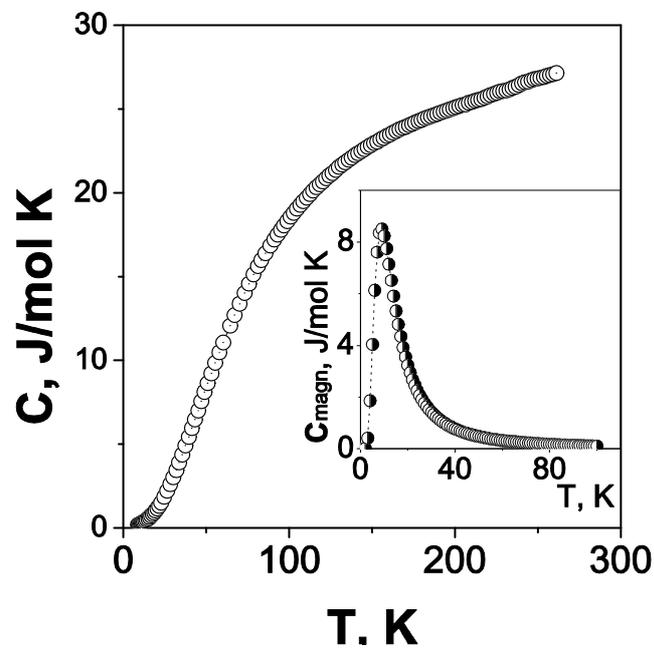

Figure 6. Specific heat of pure cpo. Inset shows the magnetic contribution $C_{magn}$ calculated according to Eq.6 (with $P = 1$)

When neglecting exchange interaction between cpo-dimers, the magnetic susceptibility is given by

$$\chi_m = \frac{PN_A g^2 \mu_B^2}{3k_B T} 6\exp(2J/k_B T)/[1+3\exp(2J/k_B T)]. \tag{5}$$

When comparing this expression with experimental data (Fig.5), we can estimate the exchange parameter as $J = -13$ K and the purity factor as $P = 0.002$ (very large degradation). This value is close to that obtained above for [Mn(hfac)$_2$(cpo)]$_2$.

The corresponding theoretical result for magnetic specific heat reads ($R$ is the universal gas constant)

$$C_{magn} = 12PR(J/k_B T)^2 \exp(2J/k_B T)/[1+3\exp(2J/k_B T)]^2. \tag{6}$$

This expression (up to the factor of 2) should describe magnetic specific heat for the complex [Mn(hfac)$_2$(cpo)]$_2$ too, since the nearly free Mn ions do not contribute to specific heat at not too low temperatures. At the same time, they can modify somewhat the exchange interaction in the cpo-dimers.

Because of smallness of the purity factor, the magnetic contribution can be hardly picked up on the background of the lattice specific heat.

**Conclusions**

In the new polynuclear complex [Mn(hfac)$_2$(cpo)]$_2$ the 4-cyanopyridine-N-oxide molecule has a singlet ground state and the corresponding exchange interactions are antiferromagnetic. The hierarchy of exchange parameters determines the non-trivial dependence of magnetic susceptibility. At the same time, owing to the branching of the chemical bonds the complex shows a ferromagnetic-like behavior with $S_{tot} = 5$ in the ground state. Although the intercluster ferromagnetic exchange interaction is very small, this is essential for stabilizing the structure of the molecular magnet.

It should be noted that a similar physics can occur in other magnetic cluster systems, e.g., in tetranuclear Mn$^{II/III}$ complexes where the Heisenberg model for two Mn$^{II}$ and two Mn$^{III}$ spins should be considered[14,15].

The results of the present work open new opportunities in searching ferromagnetic high-spin clusters, besides using high-spin radicals and complexes with ferromagnetic exchange coupling between the magnetic units. This is important in connection with the magnetic molecular design problem.

This work is supported in part by the Programs of RAS Presidium "Quantum mesoscopic and disordered structures", project No. 12-P-2-1041, and "Synthesis of functional nitroxyls for design of new magnetoactive substances and materials", project No. 12-M-23-2054
.